\documentclass{acm_proc_article-sp}

\begin{document}

\title{Quantitative Methods for Comparing Different HVAC Control Schemes\titlenote{Open source software that implements our methodology in the MATLAB language can be found at the following link \texttt{http://hybrid.eecs.berkeley.edu/\textasciitilde NEDE/compEng.zip}}}

\numberofauthors{6} %  in this sample file, there are a *total*
\author{
% 1st. author
\alignauthor Anil Aswani\\
       \affaddr{EECS, Berkeley, CA 94720}\\
       \email{aaswani@eecs.berkeley.edu}
% 2nd. author
\alignauthor Neal Master\\
       \affaddr{EECS, Berkeley, CA 94720}\\
       \email{neal.m.master@}\\
	\email{berkeley.edu}
% 3rd. author
\alignauthor Jay Taneja\\
       \affaddr{EECS, Berkeley, CA 94720}\\
       \email{taneja@cs.berkeley.edu}
\and  % use '\and' if you need 'another row' of author names
% 4th. author
\alignauthor Andrew Krioukov\\
       \affaddr{EECS, Berkeley, CA 94720}\\
       \email{kriokov@cs.berkeley.edu}
% 5th. author
\alignauthor David Culler\\
       \affaddr{EECS, Berkeley, CA 94720}\\
       \email{culler@cs.berkeley.edu}
% 6th. author
\alignauthor Claire Tomlin\\
       \affaddr{EECS, Berkeley, CA 94720}\\
       \email{tomlin@eecs.berkeley.edu}
}
% There's nothing stopping you putting the seventh, eighth, etc.
% author on the opening page (as the 'third row') but we ask,
% for aesthetic reasons that you place these 'additional authors'
% in the \additional authors block, viz.
% Just remember to make sure that the TOTAL number of authors
% is the number that will appear on the first page PLUS the
% number that will appear in the \additionalauthors section.

\maketitle
\begin{abstract}
Experimentally comparing the energy usage and comfort characteristics of different controllers in heating, ventilation, and air-conditioning (HVAC) systems is difficult because variations in weather and occupancy conditions preclude the possibility of establishing equivalent experimental conditions across the order of hours, days, and weeks. This paper is concerned with defining quantitative metrics of energy usage and occupant comfort, which can be computed and compared in a rigorous manner that is capable of determining whether differences between controllers are statistically significant in the presence of such environmental fluctuations.  Experimental case studies are presented that compare two alternative controllers (a schedule controller and a hybrid system learning-based model predictive controller) to the default controller in a building-wide HVAC system.  Lastly, we discuss how our proposed methodology may also be able to quantify the efficiency of other building automation systems.
\end{abstract}



 % NOT required for Proceedings
\section{Introduction}

Heating, ventilation, and air-conditioning (HVAC) systems contribute to a significant fraction of building energy usage.  As a result, these systems have seen an increasing amount of research towards their modeling and efficient control (e.g., \cite{nghiem2011,kelman2011,frauke2012,liao2012,aswani2011_proc,ma2012}).  The primary challenge is ensuring comparable levels of occupant comfort, in relation to existing HVAC controllers, while achieving reductions in energy consumption.  Simulations and experiments indicate that this goal is achievable for a large variety of systems.

% and environmental conditions.

Experimentally comparing the energy efficiency and comfort of different control schemes is difficult because of the large temporal variability in weather and occupancy conditions.  Furthermore, energy usage of HVAC equipment does not typically scale linearly in the relevant variables. For example, a simplistic model would state that the amount of energy $E$ required to maintain a set point of $T_s$ for the supply air temperature (SAT) of an HVAC system with a warmer outside air temperature (OAT) of $T_o (> T_s)$ is proportional to $E \sim T_o - T_s$.  We have observed on our building-wide testbed \cite{aswani_2012hvac} that such models do not capture the complex energy characteristics of the equipment.  One reason for this is that some equipment is designed to operate most efficiently at certain temperatures or settings.

There is an additional difficulty in experimentally comparing the efficiency of different control schemes.  Not all buildings have equipment to directly measure the energy consumption of only HVAC equipment.  It is common for energy measurements to include appliances, water heating, and other sources of energy consumption that are difficult to disaggregate from HVAC energy usage.  Such disaggregation is hard because the energy signature and characteristics of HVAC equipment is often similar to appliances like refrigerators or water heaters.  Another reason that disaggregation is difficult is that installing separate meters to measure energy usage of an HVAC system that is physically and electrically distributed throughout a building can be cost-prohibitive.  As a result, experimental comparison may require comparing measurements for which the absolute differences are a small percentage of the total.  Determining whether such differences are statistically significant can be challenging.

% .  The methodology we suggest is representative of the fact that buildings and conditions are highly variable, and this is reflected in requiring a different comparison for each building.

The purpose of this paper is to propose a set of quantitative methods for comparing different HVAC control schemes.  We define quantitative metrics of efficiency and comfort, and we introduce a mathematical framework for computing these metrics and then determining whether differences in these metrics are statistically significant.  For the benefit of the sustainability community, we have provided open source software that implements our comparison methodology in the MATLAB programming language.  A case study of experiments on our building-wide testbed shows the utility of the proposed techniques.

%\footnote{\texttt{http://hybrid.eecs.berkeley.edu/EDE/compEng.zip}}

\section{Existing Comparison Methods}

\begin{table}[!t]
\begin{center}
\begin{tabular}{|c|l|}
\hline
Type & Description \\
\hline
Option A & Measuring key variables that affect \\
& HVAC energy consumption \\
\hline
Option B & Directly meausuring HVAC energy \\
& consumption \\
\hline
Option C & Measuring whole building or sub- \\
& building energy consumption \\
\hline
Option D & Estimating HVAC energy consumption \\
& using simulation models \\
\hline
\end{tabular}
\end{center}
\caption{The International Performance Measurement and Verification Protocol (IPMVP) \cite{ipmvp2012} categorizes approaches to comparing energy usage into four groups.} 
\label{table:ipmvp}
\end{table}

The International Performance Measurement and Verification Protocol (IPMVP) \cite{ipmvp2012} classifies approaches to tracking energy usage into four general classes, and it is summarized in Table \ref{table:ipmvp}.  Though IPMVP is a standard protocol for evaluating energy usage of different building components, here we restrict our discussion to HVAC equipment.  One class in IPMPV is called Option A and refers to measuring key variables that affect HVAC energy consumption.  An example is measuring the OAT and the SAT settings.  Another class is called Option B and involves directly measuring energy usage of the HVAC.  Option C is the situation in which the whole building or sub-building energy usage is measured, while Option D involves utilizing simulation models that estimate HVAC energy consumption.

The approach in \cite{aswani2011_proc}, where the energy consumption of two control schemes on a single-room testbed with central air conditioning were compared, was a combination of Option B and Option D.  Mathematical models of the temperature dynamics of different control schemes and their energy characteristics were constructed in order to allow comparisons of experimentally measured HVAC energy usage to simulations over identical weather and occupancy conditions. Unfortunately, this method does not easily scale from the small system considered in \cite{aswani2011_proc} to building-sized HVAC systems because of additional complexities of building-sized systems.

%.  This is because small HVAC systems are considerably easier to model than building-sized systems. 

In \cite{siroky2011}, the energy usage of two controllers for a ceiling radiant heating system was compared using an Option A approach.  The temperature difference between the OAT and the set point for a loop of hot water was used to evaluate the energy efficiency of different control methods.  The general challenge with using Option A is that, depending on the particular HVAC equipment, the energy usage can scale in complex ways as a function of the key parameters that are measured.  Moreover, auxiliary equipment in an HVAC system (e.g., pumps, fans) often significantly contribute to energy usage; such parameters and their relationship to energy consumption are not usually measured or well understood. 

Optimization of thermal storage for campus-wide building cooling was considered in \cite{ma2012}, and the energy usage of different controllers was compared using an Option B approach coupled with a regression model of baseline performance: This indicates large energy savings when using novel controllers.  Such an analysis is only possible when direct energy measurements of the equipment are available; when this is not the case, the differences in measurements of two controllers can often be a small percentage of the total building-wide energy consumption.  A regression model alone cannot determine whether such differences are statistically significant.

% An Option A approach is likely to miss the complex energy tradeoffs due to auxiliary equipment.

We believe that a more unbiased approach for comparing HVAC is Option C combined with simple models.  Note that our proposed method also applies to Option B, when the HVAC energy usage is directly measured.  Experiments on our testbeds indicate a significant impact of OAT on the energy characteristics of HVAC systems.  Our proposed method is to use nonparametric modeling methods to compute quantitative metrics of energy usage and occupant comfort, as they relate the OAT.  This framework allows for additional nonparametric tools that can determine whether the differences in energy usage and occupant comfort of different HVAC controllers are statistically significant.

% \section{Experimental Setup for Comparisons}
\section{Experimental Setup}

In order to ensure fair comparisons between different control methods, it is imperative that the experiments be conducted using identical building and HVAC configurations.  For the sake of argument, suppose that controller 1 is the manufacturer's configuration, and controller 2 is identical to controller 1 except for that it turns the HVAC off during the night time.  Then controller 2 can achieve substantial energy savings, but this does not reflect savings due to control schemes: The energy savings are due to having different configurations of comfort levels.

When comparing HVAC controllers, settings related to occupant comfort should be kept constant because the majority of energy is used in maintaining comfort.  The specific settings that need to be kept equal will vary depending upon the building, but important variables to consider include desired temperatures in different building zones, allowable amounts of temperature deviation in these zones, and minimum and maximum amounts of air flow.  

There is another experimental issue that is subtle: It is incorrect to keep repeating the analysis as more experimental data is measured, without making corrections to the methodology.  The reason is that the probability of making errors accumulates each time the hypothesis testing methodology is used with additional data.  Corrections would need to be made to ensure that the accumulated probability of error does not grow too large.  This can be done in a principled manner using techniques from sequential hypothesis testing (e.g., \cite{wald1945,miller1981,ghosh1991}); though, we do not discuss in our paper on how to do so.  Here, we assume that the analysis is conducted once with a fixed amount of data.

% Energy savings obtained by grossly modifying such variables are due to changing the levels of comfort, and cannot be ascribed to having a better controller.

% For a fair comparison between different control methods, it is imperative that the experiments be conducted using identical building setups.  In an HVAC system, different zones of the building have thermostats that are used to set desired temperaturesIn particular, the room set points and 

% There is an additional question of ensuring a fair comparison for different controllers.  For the sake of argument, suppose that controller 1 is the manufacturer's configuration, and controller 2 is identical to controller 1 except for that it turns the HVAC off during the night time.  Then controller 2 can achieve substantial energy savings, but this does not reflect savings due to control schemes: The energy savings are due to having different configurations of comfort levels for the two controllers.  For a fair comparison, the controllers must be compared against an identical quantification of comfort.

\begin{figure}
\begin{center}
\includegraphics[clip = true, trim = 0.04in 0.10in 0.04in 0.21in]{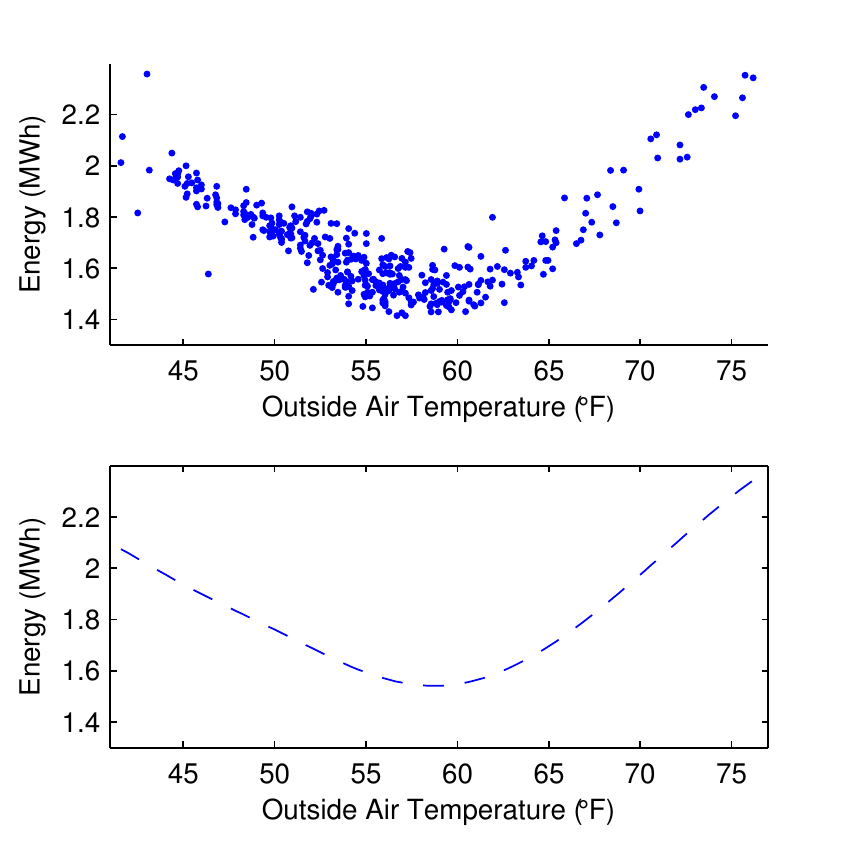}
\end{center}
\caption{\label{fig:energy}The scatter plot shows the data points $(T_0[i],E[i])$, and the smooth curve is the estimated energy characteristic $\hat{E}_{o,x}(T_0)$ for an HVAC controller.}
\end{figure}

\section{Measuring Energy Efficiency}

Without loss of generality, we assume that building-wide measurements of energy usage are available at hourly intervals.  These values will be denoted as $E[i]$ for $i = 1,\ldots,N$ measurements.  Again, note that this framework also applies to the situations where energy usage is measured at daily intervals and where the HVAC energy usage is directly measured.  Furthermore, we assume the availability of OAT measurements that correspond to the energy usage measurements: $T_o[i]$ for $i = 1,\ldots,N$.

\subsection{Quantifying Energy Consumption}

The general model describing the relationship between energy usage, OAT, occupancy $O$, and other factors (e.g., solar effects, equipment, etc.) $X$ is
\begin{equation}
E = f(T_o, O, X),
\end{equation}
where $f(\cdot,\cdot,\cdot)$ is a nonlinear relationship that is unknown.  However, occupancy and other factors are generally not directly measured (though their effects on the thermal dynamics and energy usage can be estimated using semiparametric regression \cite{aswani2011_proc,aswani_2012hvac}).  As a result, it is typically only possible to consider the energy usage averaged over occupancy and other factors
\begin{equation}
\label{eqn:energy_metric}
E_{o,x}(T_o) = \mathbb{E}\big[f(T_o, O, X) \big| T_o\big].
\end{equation}

Equation (\ref{eqn:energy_metric}) can be estimated using nonparametric regression \cite{gyorfi2002}.  (We use local linear regression in the provided code.)  It is not a single value, but is rather a curve that describes the relationship between the OAT and the average energy consumption.  Intuitively, if the data points $(T_o[i], E[i])$ for $i = 1,\ldots,N$ represent a scatter plot of energy usage versus OAT; then the estimated curve $\hat{E}_{o,x}(T_o)$ represents the smoothed version of the scatter plot.  An illustration of this can be seen in Fig. \ref{fig:energy}.

The average amount of energy used in one hour is therefore
\begin{equation}
E_{g} = \mathbb{E}_{g}\big(E_{o,x}(T_o)\big) = \mathbb{E}_{g}\big[\mathbb{E}\big[f(T_o, O, X) \big| T_o\big]\big],
\end{equation}
where $g(T_o)$ is a probability distribution of OATs.  This notation allows us to define the average amount of energy used in one day as $E_{day} = \sum_{i = 1}^{24} E_{g_i}$, where $g_i(T_o)$ is the probability distribution of OAT during the $i$-th hour of the day.  For simplicity of calculations, the source code we provide assumes a uniform distribution for the OAT.

\subsection{Interpretation of Energy Characteristics}

It is well known (e.g., \cite{kissock2002}) that the energy characteristics of HVAC have the following qualitative features: Energy usage is lowest at moderate OATs, and the consumption of energy increases as the OAT increases or decreases.  Physically, the minimum in the energy characteristic corresponds to the OAT at which the building switches between predominantly heating and predominantly cooling.  This can be seen in Fig. \ref{fig:energy}, which uses measurements taken from our building-wide HVAC testbed.  The quantitiative features of the energy characteristics vary depending upon the HVAC controller and other specific characteristics of the building and its weather microclimate.

\begin{figure}
\begin{center}
\includegraphics[clip = true, trim = 0.04in 0.02in 0.04in 0.10in]{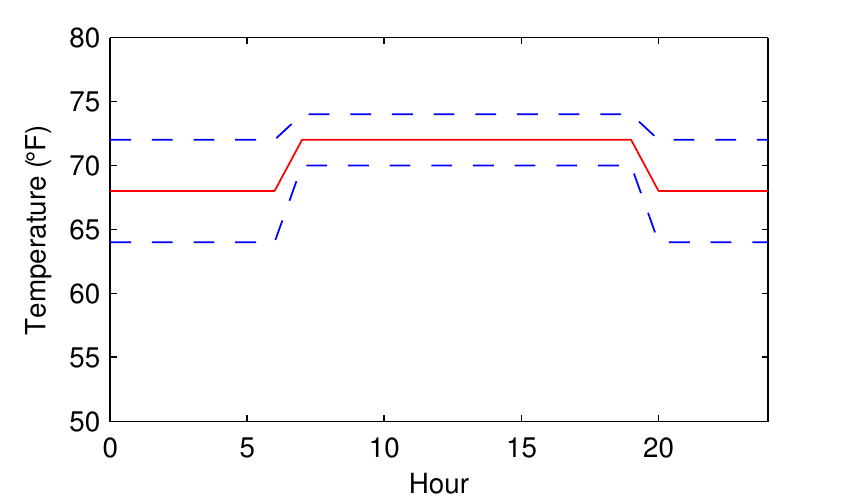}
\end{center}
\caption{\label{fig:band_of_comfort}The solid line indicates the set point of a zone throughout a single day, and the extent of the dashed lines describes the band of comfort.}
\end{figure}

\subsection{Comparing Energy Consumption}
\label{sect:hypothesis_energy}

\begin{figure}
\begin{center}
\includegraphics[clip = true, trim = 0.04in 0.10in 0.04in 0.21in]{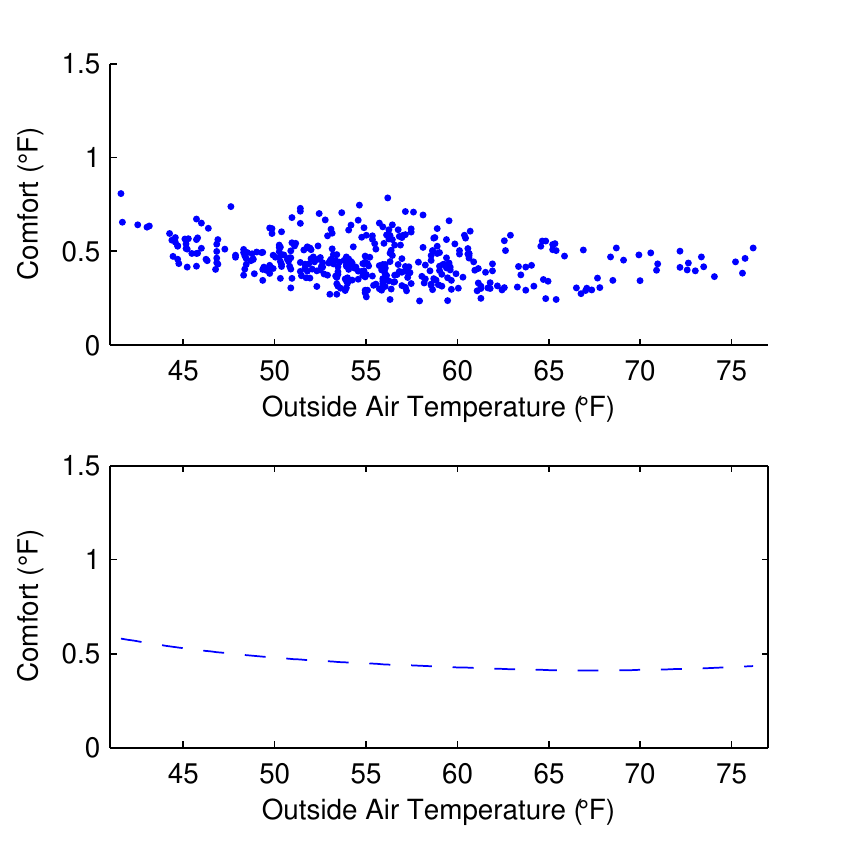}
\end{center}
\caption{\label{fig:comfort}The scatter plot shows the data points $(T_0[i],C[i])$, and the smooth curve is the estimated comfort characteristic $\hat{C}_{o,x}(T_0)$ for an HVAC controller.}
\end{figure}

We make the assumption that exactly two different control schemes are being compared; such an assumption is not restrictive, because we can do a set of pairwise comparisons with appropriate corrections made for multiple hypothesis testing.  The two controllers being compared will be referred to as controller 1 and controller 2, and they will be denoted using a superscript 1 and 2, respectively.  Though we could compute the average amount of energy used in one day for each controller, there is no guarantee \textit{a priori} that the difference of these estimated energy characteristics $\hat{E}_{o,x}^1(T_o), \hat{E}_{o,x}^2(T_o)$ is statistically significant

%quantities $\Delta \hat{E}^{2,1} = \hat{E}_{day}^2 - \hat{E}_{day}^1$ is statistically significant.

Our approach is to use a hypothesis test to quantify the evidence for the statement that the estimated characteristics for the two controllers are equal: $\hat{E}_{o,x}^1(T_o) = \hat{E}_{o,x}^2(T_o)$.  If the $p$-value of this test is less than a significance level $\alpha$, then we can say that the difference is statistically significant.  Otherwise, the difference is not statistically significant if the $p$-value is greater than $\alpha$.  The code we provide uses a significance level of $\alpha = 0.01$, and note that the intuition is that $\alpha$ gives the probability of incorrectly concluding that a difference is statistically significant.

% Because the curves $\hat{E}_{o,x}^1(T_o), \hat{E}_{o,x}^2(T_o)$ are computed using nonparametric methods (e.g., the Nadaraya-Watson estimator), it is not possible to use common methods like the $t$-test to compare them.  Nonparametric hypothesis tests based on bootstrap procedures are an attractive alternative for our situation \cite{boostrap2007}.  However, there is a temporal correlation between the measured energy usage $E[i]$ for different values of $i$.  This precludes the use of resampling residuals boostrap, which is a standard bootstrap method.

Because the curves $\hat{E}_{o,x}^1(T_o), \hat{E}_{o,x}^2(T_o)$ are computed using nonparametric methods, it is not possible to use common methods like the $t$-test to compare them.  Nonparametric hypothesis tests based on bootstrap procedures are an attractive alternative for our situation \cite{boostrap2007}.  However, there is a temporal correlation between the measured energy usage $E[i]$ for different values of $i$.  This precludes the use of resampling residuals bootstrap, which is a standard bootstrap method.

The methodology we propose makes use of a nonparametric hypothesis test that utilizes the moving block bootstrap.  This form of bootstrap is designed to handle dependent data in which there is a temporal correlation between measurements.  The interested reader can refer to \cite{boostrap2007} for details about this hypothesis test, and this is the technique that is implemented in our code.  This is used to determine whether $\hat{E}_{o,x}^1(T_o), \hat{E}_{o,x}^2(T_o)$ are statistically different.  We can also use the same methodology to determine whether the difference in the estimated average energy usage over one day $\Delta \hat{E}^{2,1} = \hat{E}_{day}^2 - \hat{E}_{day}^1$ is not equal to zero to a statistically significant level.  Note that because two hypotheses are being tested, we must make appropriate adjustments \cite{miller1981}: Our code uses a Bonferroni correction to generate adjusted $p$-values \cite{wright1992}.  

\subsection{Confidence Intervals of Energy Consumption} 
\label{sect:confidence_energy}

If the hypothesis test in Sect. \ref{sect:hypothesis_energy} indicates that $\Delta \hat{E}^{2,1} \neq 0$ to a statistically significant level, then we can interpret both the value and magnitude of this quantity.  If $\Delta \hat{E}^{2,1} < 0$, then this means that controller 2 uses less energy than controller 1 on average, and the absolute value $|\Delta \hat{E}^{2,1}|$ is the average amount of energy savings over a day due to controller 2 as compared to controller 1.  The opposite statement holds if $\Delta \hat{E}^{2,1} > 0$.  

% have associated levels of uncertainty
Since $\Delta \hat{E}^{2,1}$ is an estimated quantity, it will itself have uncertainty.  In order to better characterize the difference between the energy consumption of controllers 1 and 2, it is useful to also compute a confidence interval for $\Delta \hat{E}^{2,1}$.  For a confidence level of $\beta \in (0,1]$, the confidence interval contains the true quantity (e.g., $\Delta E^{2,1}$) for $\beta$ fraction of the experiments.  The same moving block bootstrap methodology used in Sect. \ref{sect:hypothesis_energy} can be used to estimated a bias-corrected bootstrap confidence interval \cite{efron1981}.

\section{Measuring Occupant Comfort}

\begin{figure}
\begin{center}
\includegraphics[width=3.3in]{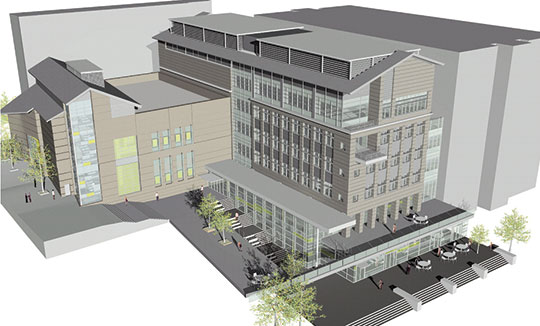}
\end{center}
\caption{\label{fig:sdh}Sutardja Dai Hall is 141,000-square-foot building, and our BRITE-S testbed is within this building.}
\end{figure}

\begin{table*}
\begin{center}
\begin{tabular}{|l|l|c|l|c|}
\hline
Controller 2 & Energy Characteristics & $\Delta \hat{E}^{2,1}$ & Comfort Characteristics &  $\Delta \hat{C}^{2,1}$  \\
\hline
Temperature and & Statistically Different & $-1.16$MWh & Statistically Different & $0.35^\circ F$ \\
Reheat Schedule &  ($p = 0.002$) & ($p = 0.08$) & ($p = 0.002$) & ($p = 0.9$) \\
\hline
LBMPC SAT & Statistically Different & $-1.53$MWh & \textbf{NOT} Statistically Different & $-0.75^\circ F$ \\
Control & ($p = 0.002$) & ($p = 0.002$) & ($p = 0.8$) & $(p = 0.5$) \\
\hline
\end{tabular}
\end{center}
\caption{Summary of case studies on the BRITE-S testbed.  Controller 1 is the default, manufacturer-provided controller, and controller 2 is as indicated.  Also, $\alpha = 0.05$ was taken to be the significance level of the hypothesis tests.}
\label{table:summary}
\end{table*}

Quantifying the comfort levels of occupants is difficult because it is a function of many variables: metabolic rate, clothing insulation, air temperature, radiant temperature, air speed, and humidity \cite{ashrae55}.  This problem is further complicated by the fact that most buildings are not instrumented to measure these variables.  

In order to define a quantification of comfort that is both tractable and will scale to many buildings, we focus on a measure that is only dependent on temperature; temperature measurements are usually available through existing thermostats.  We assume that the average temperature for each of the $j = 1,\ldots,Z$ zones are measured at hourly intervals $T_j[i]$ for $i = 1,\ldots,N$.  

\subsection{Band of Comfort}

The $j$-th zone of a building will have a temperature set point $S_j$ that can vary throughout the day, and we assume that we have recorded the average value of this set point at hourly intervals: $S_j[i]$ for $i = 1, \ldots, N$.  Furthermore, we assume that comfort is defined as maintaining zone temperature within $\pm B_j[i]$ for $i = 1, \ldots, N$ of the set point.  It can be visualized as a band of comfort, as shown in Fig. \ref{fig:band_of_comfort}.  This band is allowed to vary because the configured comfort levels often depend on the time of day or week; an example is allowing greater temperature fluctuations in an office building at night.

Our justification for this notion of comfort is that it corresponds to the Class A, Class B, and Class C notions of comfort as defined in \cite{ashrae55}.  Furthermore, this quantification is implicitly making the assumption that the set point indicates the true occupant preference.  It represents a compromise between tractability of quantifying comfort and maintaining a reasonable metric.

\subsection{Quantifying Occupant Comfort}

Let $(x)_+$ be the thresholding function, which is defined so that $(x)_+ = 0$ if $x < 0$ and $(x)_+ = x$ otherwise.  We define our quantification of comfort using soft thresholding as
\begin{equation}
C = 1/Z \cdot \textstyle\sum_{j=1}^Z \textstyle\int_0^1 (|T_j - S_j|-B_j)_+ dt,
\end{equation}
where the integral with respect to $dt$ is over one hour of time.  The intuition is that this quantity increases whenever a zone temperature leaves the band of comfort, and the amount of increase in this quantity is proportional to the amount and duration of temperature deviation.

Because $T_j$ is implicitly a function of the outside temperature $T_0$, occupancy $O$, and other factors $X$, we can abstractly represent the comfort quantification as 
\begin{equation}
C = h(T_0, O, X),
\end{equation}
where $h(\cdot,\cdot,\cdot)$ is a nonlinear relationship that is unknown.  Just as in the case of energy, we consider the comfort averaged over occupancy and other factors
\begin{equation}
C_{o,x}(T_0) = \mathbb{E}\big[h(T_0, O, X)\big| T_0\big],
\end{equation}
because they are not measured.  We can again generate a scatter plot of the data points $(T_o[i],C[i])$ for $i = 1,\ldots,N$, where $C[i] = \sum_j (|T_j[i] - S_j[i]| - B_j[i])_+$.  And a smoothed version can be estimated $\hat{C}_{o,x}(T_0)$ using nonparametric regression.  An example of this is seen in Fig. \ref{fig:comfort}.  

The average occupant comfort over one hour is given by
\begin{equation}
C_g = \mathbb{E}_g\big(C_{o,x}(T_0)\big) = \mathbb{E}_g\big[\mathbb{E}\big[h(T_0, O, X)\big|T_o\big]\big],
\end{equation}
where $g(T_o)$ is a probability distribution of OATs.  We can then define the average amount of comfort in one day as $C_{day} = \sum_{i = 1}^{24} C_{g_i}$, where $g_i(T_o)$ is the probability distribution of OAT during the $i$-th hour of the day.  Our source code assumes a uniform distribution for the OAT.

\subsection{Interpretation of Comfort Characteristics}

The purpose of an HVAC system is to provide uniform levels of occupant comfort regardless of external and internal conditions, and so the ideal HVAC controller will provide a comfort characteristic that is constant across all OATs. However, in practice it is more difficult for the HVAC to maintain the building environment when the weather is more extreme.  This leads to slight decreases in comfort (i.e., the comfort characteristic is higher) at low and high OATs.  This can be seen in Fig. \ref{fig:comfort}, which uses measurements taken from our building-wide HVAC testbed.  The quantitative features of the comfort characteristics vary depending upon the HVAC controller and other specific characteristics of the building and its weather microclimate.

\subsection{Comparing Occupant Comfort}

Using the hypothesis testing methodology outlined in Sect. \ref{sect:hypothesis_energy}, we can determine (a) whether the estimated comfort characteristics $\hat{C}_{o,x}^1(T_o), \hat{C}_{o,x}^2(T_o)$ for the two controllers are statistically different, and (b) whether the difference in the estimated average comfort over one day $\Delta \hat{C}^{2,1} = \hat{C}_{day}^2 - \hat{C}_{day}^1$ is not equal to zero to a statistically significant level.  Like in Sect. \ref{sect:hypothesis_energy}, we must correct for multiple comparison effects. 

\subsection{Confidence Intervals for Occupant Comfort}

If the hypothesis test in Sect. \ref{sect:hypothesis_energy} indicates that $\Delta \hat{C}^{2,1} \neq 0$ to a statistically significant level, then we can interpret both the value and magnitude of this quantity.  Because lower values of $C$ correspond to greater comfort levels, we can interpret the quantity $\Delta \hat{C}^{2,1}$ in the same way as $\Delta \hat{E}^{2,1}$ is interpreted in Sect. \ref{sect:confidence_energy}. And because $\Delta \hat{C}^{2,1}$ is an estimated quantity, the difference between the comfort levels of controllers 1 and 2 can be better characterized by also computing its confidence interval as in Sect. \ref{sect:confidence_energy}.

\section{Case Study: BRITE-S Testbed}

The Berkeley Retrofitted and Inexpensive HVAC Testbed for Energy Efficiency in Sutardja Dai Hall (BRITE-S) platform \cite{krioukov2011,aswani2012_brites} is a building-wide HVAC system that maintains the indoor environment of a 141,000-square-foot building, shown in Fig. \ref{fig:sdh}, that is divided between a four-floor nanofabrication laboratory (NanoLab) and seven floors of general space (including office space, classrooms, and a coffee shop).  The building automation equipment can be measured and actuated through a BACnet protocol interface.

The HVAC system uses a 650-ton chiller to cool water.  Air-handler units (AHUs) with variable-frequency drive fans distribute air cooled by the water to variable air volume (VAV) boxes throughout the building.  Since the NanoLab must operate within tight tolerances, our control design can only modify the operation of the general space AHUs and VAV boxes, with no modification of chiller settings that are shared between the NanoLab and general space.

The default, manufacturer-provided controller in BRITE-S uses PID loops to actuate the VAV boxes and keeps a constant SAT within the AHUs. Conventional SAT reset control is not possible because several VAV boxes for zones with computer equipment provide maximum air flow rates throughout the entire day for all, except the coldest SATs. 

Here, we present case studies that compare this default controller (controller 1) with two different controllers.  Controller 2 will refer to the controller being compared to controller 1.  Note that the building is configured to provide a band of comfort of $\pm B_j[i] = \pm 1^\circ$F for all times and zones.  The results of these case studies are summarized in Table \ref{table:summary}, and details are given below.

\subsection{Temperature and Reheat Schedule}

\begin{figure}
\begin{center}
\includegraphics[clip = true, trim = 0.04in 0.10in 0.04in 0.21in]{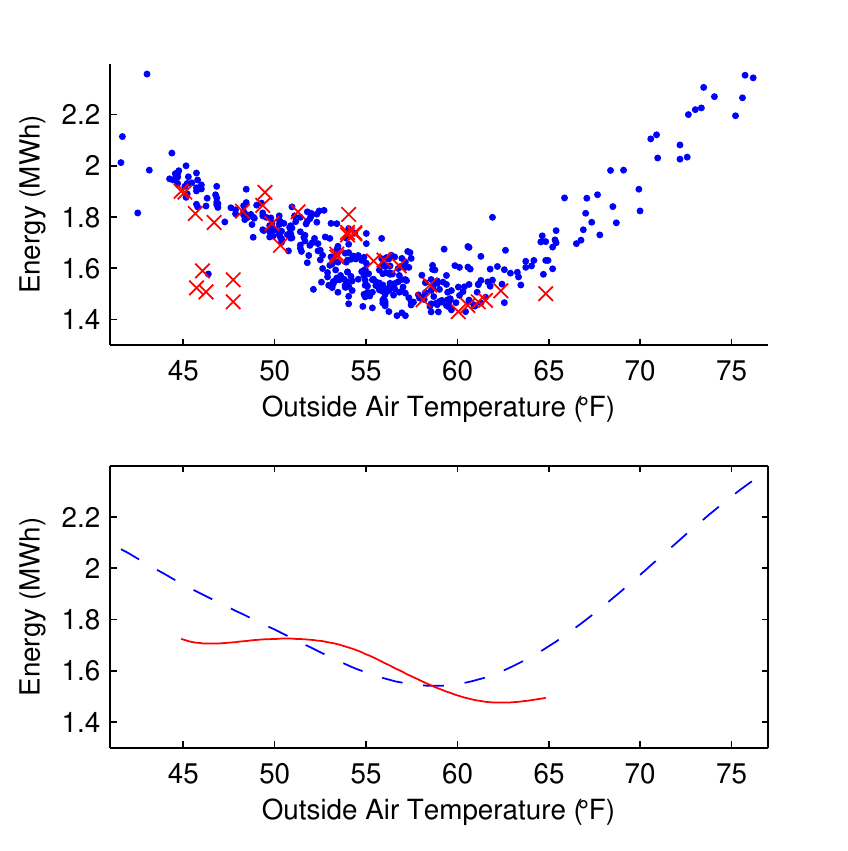}
\end{center}
\caption{\label{fig:schedule_energy}
The cross marks and solid line (points and dashed line) denote the energy characteristics of the schedule (default) controller.}
% \caption{The cross marks and solid line denote the energy usage of the schedule controller, while the dot marks and dashed line denote the energy usage of the default controller.}
\end{figure}

%This controller turns off all heating in the building during the night time, and it matches the SAT to the OAT in an effort to reduce the amount of energy used to cool the air.  According to an Option A analysis that neglects energy usage of auxiliary equipment like fans, this control will use nearly zero energy.  However, our comparison methodology shows that this is not the case.

The first controller to be compared to controller 1 turns off all heating in the building during the night time, and it matches the SAT to the OAT in an effort to reduce the amount of energy used to cool the air.  We call this the schedule controller.  An Option A analysis that neglects auxiliary equipment like fans will estimate nearly zero energy usage.  However, our comparison methodology shows that this is not correct.

% The estimated energy characteristics of the default controller $\hat{E}_{o,x}^1(T_0)$ and the schedule controller $\hat{E}_{o,x}^2(T_0)$ are seen in Fig. \ref{fig:schedule_energy}, and the difference between the energy characeteristics is statistically significant $(p = 0.001)$.  However, the estimated difference in average energy usage over one day $\Delta \hat{E}^{2,1} = -1.16$MWh is not statistically significant $(p=0.6)$.  This means that, statistically speaking, there is not enough evidence to conclude that the true difference is not zero (i.e., $\Delta E^{2,1} = 0$MWh).

The estimated energy characteristics $\hat{E}_{o,x}^1(T_0),\hat{E}_{o,x}^2(T_0)$ are shown in Fig. \ref{fig:schedule_energy}, and their differences are statistically significant $(p = 0.002)$.  However, the estimated difference in average energy usage over one day $\Delta \hat{E}^{2,1} = -1.16$MWh is not statistically significant $(p=0.08)$, meaning that there is not enough evidence to exclude that $\Delta E^{2,1} = 0$MWh.

% The estimated comfort characteristics of the default controller $\hat{C}_{o,x}^1(T_0)$ and the schedule controller $\hat{C}_{o,x}^2(T_0)$ are seen in Fig. \ref{fig:schedule_comfort}, and the difference between the comfort characeteristics is statistically significant $(p = 0.001)$.  However, the estimated difference in average comfort over one day $\Delta \hat{C}^{2,1} = 0.35^\circ$F is not statistically significant $(p=0.8)$.  This means that, statistically speaking, there is not enough evidence to conclude that the true difference is not zero (i.e., $\Delta C^{2,1} = 0^\circ$F).

\begin{figure}
\begin{center}
\includegraphics[clip = true, trim = 0.04in 0.10in 0.04in 0.21in]{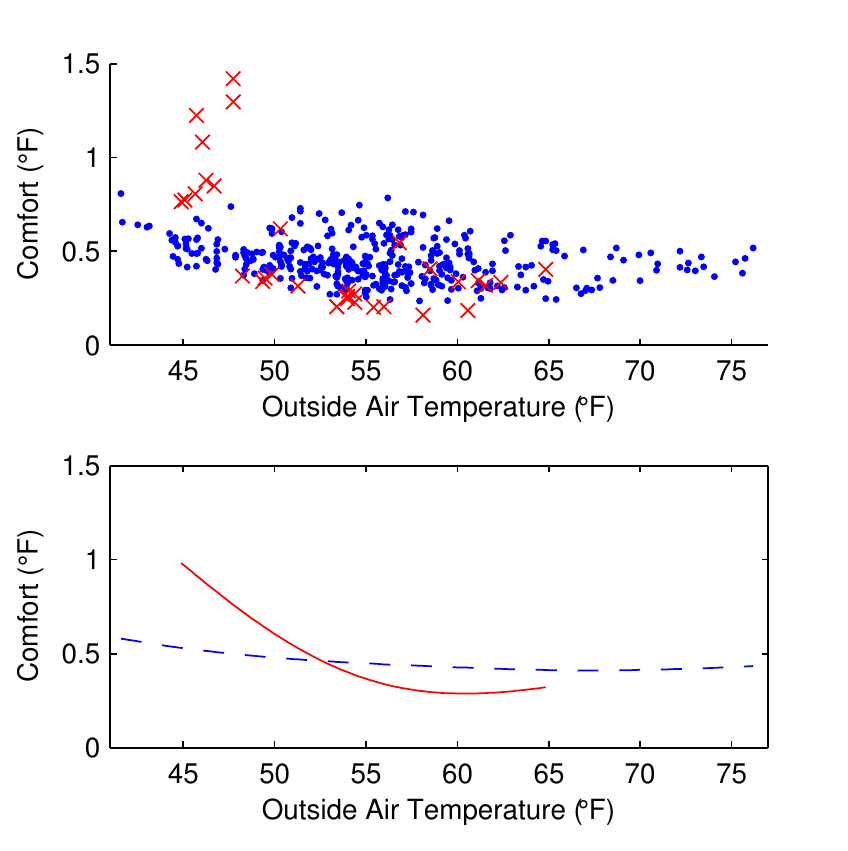}
\end{center}
\caption{\label{fig:schedule_comfort}
The cross marks and solid line (points and dashed line) denote the comfort characteristics of the schedule (default) controller.}
\end{figure}

The estimated comfort characteristics $\hat{C}_{o,x}^1(T_0),\hat{C}_{o,x}^2(T_0)$ are shown in Fig. \ref{fig:schedule_comfort}, and their differences are statistically significant $(p = 0.002)$.  However, the estimated difference in average comfort over one day $\Delta \hat{C}^{2,1} = 0.35^\circ$F is not statistically significant $(p=0.9)$, meaning that there is not enough evidence to exclude that $\Delta C^{2,1} = 0^\circ$F.

% The estimated characteristics show some qualitative differences between the two controllers.  

Though the schedule controller saves energy by turning off heating at night, these savings are somewhat negated by having to heat the building to a comfortable temperature during the day (i.e., the hump in $\hat{E}_{o,x}^2(T_0)$ at $55^\circ$F).  As compared to the default controller, the schedule controller substantially degrades comfort at colder temperatures and provides slight improvements in comfort at moderate temperatures.

\subsection{LBMPC Control of Supply Air Temperature}

The second controller to be compared to controller 1 uses a hybrid system learning-based model predictive controller \cite{aswani2011_safe,aswani2012_brites} to determine a sequence of SATs amongst the values of $52^\circ$F, $58^\circ$F, and $64^\circ$F.  An Option A analysis provides overly optimistic estimates of the energy savings due to this control method, while the methodology described in this paper shows modest energy savings. % and occupant comfort.

% The estimated energy characteristics of the default controller $\hat{E}_{o,x}^1(T_0)$ and the LBMPC controller $\hat{E}_{o,x}^2(T_0)$ are seen in Fig. \ref{fig:lbmpc_energy}, and the difference between the energy characeteristics is statistically significant $(p = 0.001)$.  The estimated difference in average energy usage over one day $\Delta \hat{E}^{2,1} = -1.53$MWh is statistically significant $(p=0.006)$.  And the 95\% confidence interval is $\Delta \hat{E}^{2,1} \in [-2.11,-1.02]$MWh.

\begin{figure}
\begin{center}
\includegraphics[clip = true, trim = 0.04in 0.10in 0.04in 0.21in]{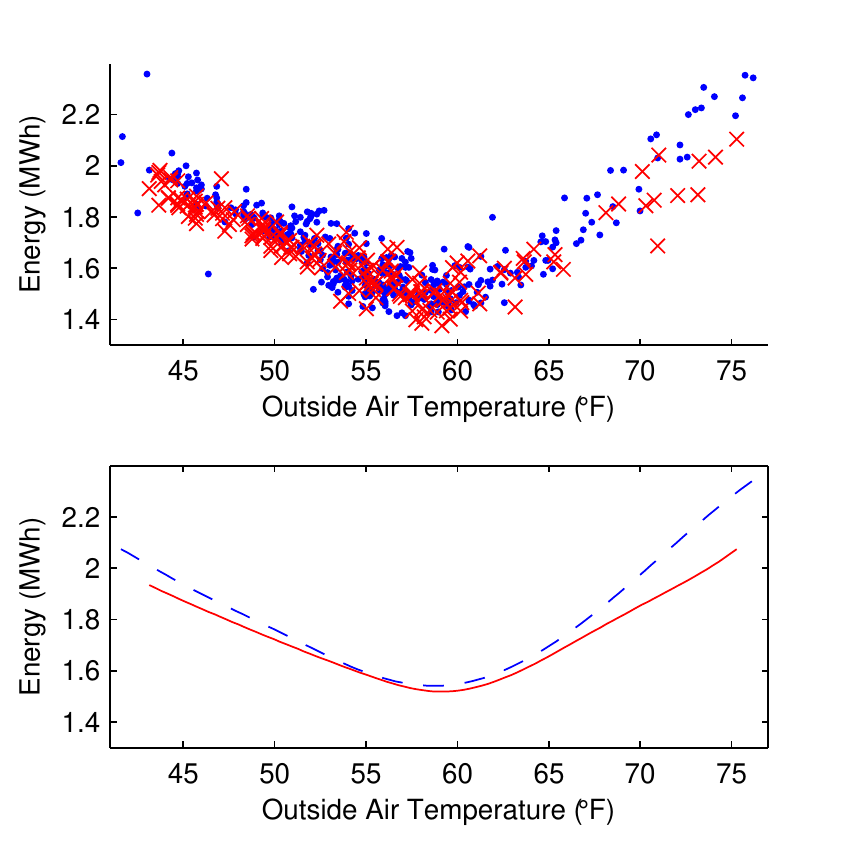}
\end{center}
\caption{\label{fig:lbmpc_energy}The cross marks and solid line (points and dashed line) denote the energy characteristics of the LBMPC (default) controller.}
\end{figure}

The estimated energy characteristics $\hat{E}_{o,x}^1(T_0),\hat{E}_{o,x}^2(T_0)$ are shown in Fig. \ref{fig:lbmpc_energy}, and their differences are statistically significant $(p = 0.002)$.  The estimated difference in average energy usage over one day $\Delta \hat{E}^{2,1} = -1.53$MWh is statistically significant $(p=0.002)$.  And the 95\% confidence interval is $\Delta \hat{E}^{2,1} \in [-2.07,-1.02]$MWh.

% The estimated comfort characteristics of the default controller $\hat{C}_{o,x}^1(T_0)$ and the LBMPC controller $\hat{C}_{o,x}^2(T_0)$ are seen in Fig. \ref{fig:lbmpc_comfort}, and the difference between the comfort characeteristics is not statistically significant $(p = 0.4)$.  Furthermore, the estimated difference in average comfort over one day $\Delta \hat{C}^{2,1} = -0.75^\circ$F is not statistically significant $(p=0.3)$.  This means that, statistically speaking, there is not enough evidence to conclude that the true difference is not zero (i.e., $\Delta C^{2,1} = 0^\circ$F).

\begin{figure}
\begin{center}
\includegraphics[clip = true, trim = 0.04in 0.10in 0.04in 0.21in]{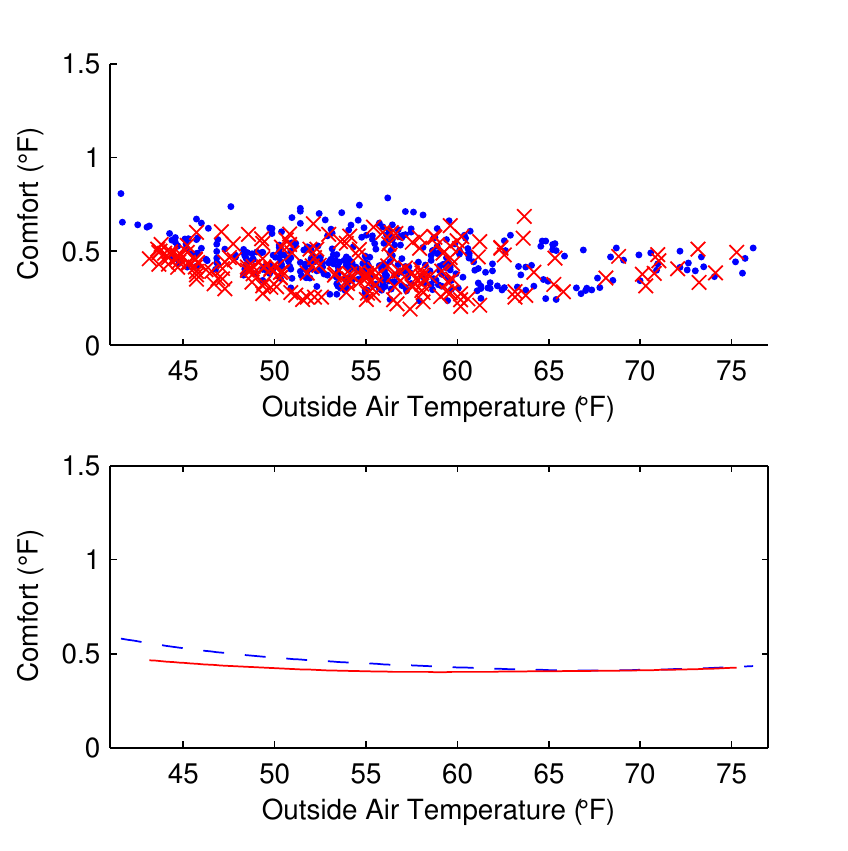}
\end{center}
\caption{\label{fig:lbmpc_comfort}The cross marks and solid line (points and dashed line) denote the comfort characteristics of the LBMPC (default) controller.}
\end{figure}

The estimated comfort characteristics $\hat{C}_{o,x}^1(T_0),\hat{C}_{o,x}^2(T_0)$ are shown in Fig. \ref{fig:lbmpc_comfort}, and their differences are not statistically significant $(p = 0.8)$.  Furthermore, the estimated difference in average comfort over one day $\Delta \hat{C}^{2,1} = -0.75^\circ$F is not statistically significant $(p=0.5)$, meaning that there is not enough evidence to exclude that $\Delta C^{2,1} = 0^\circ$F.

The LBMPC controller provides modest energy savings at most OATs, which sum up to significant savings over a day.  And because the difference in comfort characteristics is not statistically significant, this suggests that the LBMPC and default controllers provide comparable levels of comfort.

\section{Conclusion}

We have presented quantitative metrics for comparing the energy usage and comfort of different HVAC controllers.  These metrics can be computed relatively easily for a wide variety of buildings and are amenable to methods that can determine whether experimental differences between controllers are statistically significant.  Though our focus here has been on energy-efficient HVAC control, this methodology can be used to compare the efficiency of different building components such as water heating, lighting, and even changes in the comfort level of HVAC.

\section{Acknowledgments}
The authors thank Ram Vasudevan for his insightful suggestions and comments, in addition to support from Domenico Caramagno who has graciously allowed experiments on the building he manages.  This material is based upon work supported by the National Science Foundation under Grant CNS-0931843 (CPS-ActionWebs) and CNS-0932209 (CPS-LoCal). The views and conclusions contained in this document are those of the authors and should not be interpreted as representing the official policies, either expressed or implied, of the National Science Foundation.

\bibliographystyle{abbrv}
\bibliography{ifacconf}  % sigproc.bib is the name of the Bibliography in this case
% ACM needs 'a single self-contained file'!
%\balancecolumns

\balancecolumns
% That's all folks!
\end{document}